\documentclass{biophys-new}
\usepackage[utf8]{inputenc}
\usepackage{graphicx}
\usepackage[colorlinks,allcolors=cyan!70!black]{hyperref}

\usepackage{lipsum}

\title{Heterogeneity in ER network structure guides diffusive transport and kinetics}
\runningtitle{ER heterogeneity} 

\author[1]{Zubenelgenubi C. Scott}
\author[2]{Katherine Koning}
\author[2] {Molly Vanderwerp}
\author[3] {Lorna Cohen}
\author[2]{Laura M. Westrate}
\author[1,*]{Elena F. Koslover}
\runningauthor{Scott et al.} 

\affil[1]{Department of Physics, University of California, San Diego, La Jolla, CA 92093, USA}
\affil[2]{Department of Chemistry and Biochemistry, Calvin University, Grand Rapids, MI 49546, USA}
\affil[3]{Van Andel Institute, Grand Rapids MI 49503}

\corrauthor[*]{ekoslover@uscd.edu}

\papertype{Article}

\newcommand{\needcite}{{\bf[CITE]} }

\graphicspath{{./figures/}}

\begin{document}

\begin{frontmatter}

\begin{abstract}


The endoplasmic reticulum (ER) is a dynamic network of interconnected sheets and tubules that orchestrates the distribution of lipids, ions, and proteins throughout the cell. The impact of its complex, dynamic morphology on its function as an intracellular transport hub remains poorly understood.
To elucidate the functional consequences of ER network structure and dynamics, we quantify how the heterogeneity of the peripheral ER in COS7 cells affects diffusive protein transport.
 {\em In vivo} imaging of photoactivated ER membrane proteins demonstrates their non-uniform spreading to adjacent regions, in a manner consistent with simulations of diffusing particles on  extracted network structures. Using a minimal network model to represent tubule rearrangements, we demonstrate that ER network dynamics are sufficiently slow to have little effect on diffusive protein transport. 
  Furthermore, stochastic simulations reveal a novel consequence of ER network heterogeneity: the existence of 'hot spots' where sparse diffusive reactants are more likely to find one another. Intriguingly, ER exit sites
are disproportionately found in these highly accessible regions.
Combining {\em in vivo} experiments with analytic calculations, quantitative image analysis, and computational modeling, we demonstrate how structure guides diffusive protein transport and reactions in the ER.

\end{abstract}

\begin{sigstatement}
The endoplasmic reticulum (ER) is the largest organelle in the eukaryotic cell, forming a web of interconnected hollow tubules and sheets. The ER is central to the transport of many cellular components such as lipids, ions, and proteins. However, the impact of the ER's complex network architecture on these transport processes remains opaque. Using live-cell experiments and simulations, we demonstrate that structural heterogeneity leads to non-uniform transport of proteins to nearby regions of the ER. As a consequence, certain regions of the network function as `hot spots' where diffusive reactants are more likely to find each other. In live cells, sites of protein export are preferentially localized to these regions.
\end{sigstatement}
\end{frontmatter}


\section*{Introduction}

The eukaryotic cell contains a myriad of complex structures and compartments, each serving a specialized functional role. These include the tortuous interior of interconnected mitochondria~\cite{dieteren2011solute}, the stacked sheets~\cite{terasaki2013stacked} and tubular networks~\cite{wang2016cooperation,nixon2016increased} of the perinuclear and peripheral endoplasmic reticulum (ER), and the intertwined actin and microtubule networks of the cytoskeleton~\cite{burute2019cellular,ross2008cargo,mogre2020getting}.
The morphology of these intracellular structures modulates the long-range active and passive transport of particles within them~\cite{agrawal2022morphology}.
For example, the winding cristae of mitochondria slow down the long-range spread of particles~\cite{dieteren2011solute}, while spiral dislocations connecting ER sheets facilitate more rapid diffusive transport~\cite{terasaki2013stacked,huber2019terasaki}.

A number of theoretical studies have demonstrated that the architecture of the domain can play an important role in determining reaction rates, a general phenomenon described as `geometry-controlled kinetics'~\cite{benichou2010geometry}. 
%
 %
 Emergent kinetic behaviors such as ultrasensitivity, bistability, and proofreading can be promoted or suppressed when enzyme and reactant diffusion is perturbed by crowding or by association with cellular structures~\cite{takahashi2010spatio,abel2012membrane,galstyan2020proofreading}. Additional effects arise when the domain structure is dynamic, leading to time-varying effective diffusivity~\cite{chubynsky2014diffusing,jain2016diffusing} and broadening the distribution of search times~\cite{lanoiselee2018diffusion}.


One important class of intracellular geometries includes network structures, consisting of effectively one-dimensional edges connected at junction nodes.
The transport properties of spatial networks~\cite{barthelemy2011spatial} have been studied in a variety of contexts, from porous media~\cite{bryant1993physically}, to neuronal maintenance~\cite{williams2016dendritic,sartori2020statistical}.  
 For instance, particles diffusing through networks of tubes and containers have been shown to exhibit novel transport properties such as wavelike concentration fluctuations~\cite{lizana2005diffusive}, as well as enhanced reaction rates~\cite{li2014calculated}.
 
The peripheral ER~\cite{lin2017modeling, sun2022unraveling,tikhomirova2022role} and the mitochondrial networks of yeast and mammalian cells~\cite{viana2020mitochondrial,zamponi2018mitochondrial} can both be described as spatial networks of interconnected hollow tubules.
Studies of search kinetics in these networks have highlighted the importance of network connectivity, as described by the number of loops within the network~\cite{viana2020mitochondrial,brown2020impact}. The connectivity can be biologically perturbed by mutations in ER morphogens~\cite{wang2016cooperation,tikhomirova2022role} and mitochondrial fusion and fission proteins~\cite{viana2020mitochondrial,zamponi2018mitochondrial}. Prior studies have focused largely on global network architecture and transport properties, such as mean first-passage times averaged over the entire network.  Cellular networks, however, are not homogeneous lattices, implying that a significant amount of variability should be expected in local transport to specific regions~\cite{scott2021diffusive}. This variability has the potential to modulate encounter kinetics and dispersal to different regions of the cell.
 

The dynamic, interconnected web of the ER plays an important biological role as a delivery network for proteins, ions, and lipids throughout the cell~\cite{konno2021endoplasmic,yang2023diffusive,westrate2015form,wu2018here}. For example, phospholipids manufactured in the ER, must diffuse through its membrane to contact sites with lipid droplets, mitochondria, and other organelles in order to be transferred to their eventual cellular destinations~\cite{balla2019lipid,wu2018here}. Additionally, alteration of network structure through modulating expression of ER morphogens has been shown to affect the magnitude of calcium release, possibly due to altered transport through the ER lumen~\cite{konno2021endoplasmic}.

The ER also serves as a quality-control hub for newly synthesized proteins destined for secretion~\cite{ellgaard2003quality,phillips2020protein}. These proteins are co- or post-translationally inserted into the ER lumen or membrane, interact with a variety of ER-resident chaperones to ensure correct folding, and exit the organelle after encountering an ER exit site (ERES). The ERES are punctate, persistent structures which package secretory cargo into COPII-coated vesicles for subsequent transport to the Golgi apparatus~\cite{barlowe1994copii,matsuoka1998copii,watson2006sec16,westrate2020vesicular}. While in the ER, the proteins engage in diffusive transport to encounter their chaperone binding partners and to find the exit sites. Furthermore, certain steps in the protein quality control pathways are thought to occur in specialized local regions of the ER~\cite{kamhi2001novel,sontag2014sorting}, necessitating transport of proteins into and out of these regions. Given that many of the biological functions of the ER rely on its ability to serve as a topologically isolated transport network throughout the cell, understanding how network architecture modulates particle transport and encounter kinetics forms an important problem in cell biology.

In this work, we focus on the spatial heterogeneity of the peripheral ER network in mammalian (COS7) cells. We demonstrate that structural variability across individual ER networks translates to heterogeneous diffusive accessibility for different ER regions within the same cell. 
Live-cell imaging data is used to show that locally photoactivated membrane proteins spread non-uniformly to nearby regions of the ER, in agreement with simulation results that predict preferential transport to better-connected regions of the network. The contribution of dynamic ER network rearrangements is quantified using a minimal network model~\cite{lin2014structure,lin2017modeling}, and shown to have little effect on membrane protein spreading.
Furthermore, with the aid of stochastic simulations we demonstrate that the heterogeneity of the ER leads to the formation of `hot spots' where diffusing reactants are more likely to find each other, and show that ER exit sites appear to be preferentially localized to such spots in the network.
By examining the impact of ER network heterogeneity on diffusion-limited reactions and local protein spread, this work sheds light on the structure-function relationship of a biologically crucial organelle.

\section*{Materials and Methods}

\subsection*{DNA plasmids}
ER plasmids (mCherry\_KDEL, KDEL\_Venus or BFP\_KDEL) were described previously~\cite{english2013rab10,friedman2011er,zurek2011reticulon}.
Plasmids expressing fluorescently tagged COPII proteins: GFP\_Sec16s, GFP\_Sec23A, GFP\_Sec24D, and EYFP\_Sec31A were acquired from addgene (gifts from Benjamin Glick \#15775, David Stephens \#66609 and \#66613, Henry Lester \#32678)~\cite{bhattacharyya2007two,stephens2000copi,richards2011trafficking}. Generation of PAGFP\_Calnexin was performed by PCR amplification of calnexin from mEmerald\_Calnexin (gift from Michael Davidson, addgene \#54021) using iProof high-fidelity DNA Master mix (Bio-Rad) and primers flanked with Xho1 or BamH1 recognition sites (Primer Fwd: \verb=5'-AGATCTCGAGCTCATGGAAGGGAAGTGGTTGCTG -3'= and Primer Rev: \verb=5'-CCGATGGATCCCGCTCTCTTCGTGGCTTTCTGTTTCT-3'=) according to manufacturer instructions. Amplified DNA was purified using the Monarch Gel Extraction Kit (New England Bioscience) according to manufacturer protocol and digested with Xho1 and BamH1 (NEB). The digested calnexin was then ligated into PAGFP\_N1 (gift from Jennifer Lippincott-Schwartz, addgene \#11909)~\cite{patterson2002photoactivatable} using T4 DNA ligase (NEB) according to manufacture protocols. Bacterial clones were screened for insertion of calnexin sequence and confirmed by sequencing. 

\subsection*{Photoactivation experiment}

COS7 cells were purchased from ATCC and cultured in Dulbecco's modified Eagle's medium (DMEM) supplemented with 10\% fetal bovine serum (FBS) and 1\% penicillin/streptomycin. For all imaging experiments, COS7 cells were seeded in six-well, plastic-bottom dishes at $ 7.5 \times 10^{4}$ cells/ml about 16 hours before transfection. Plasmid transfections were performed using lipofectamine 3000, as described previously~\cite{hoyer2018novel}. The following standard DNA amounts were transfected per mL: 0.2 $\mu$g mCherry\_KDEL, 0.2 $\mu$g BFP\_KDEL and $0.4\mu$g PAGFP\_Calnexin. Cells were transferred to 35mm imaging dishes (CellVis) at least 16 hours before imaging.

All photoactivation experiments were performed at the Van Andel Institute Optical Microscopy Core on a Zeiss LSM 880, equipped with an Axio Observer 7 inverted microscope body, stage surround incubation, Airyscan detector, two liquid cooled MA\_PMT confocal detectors and one 32-channel GaAsP array confocal detector. Images were acquired with a Plan-Apochromat 63x (NA 1.4) oil objective with 3x optical zoom using  Zeiss Zen 2.3 Black Edition software. Photoactivated target ROIs (60x60 pixel ROI) in the peripheral ER network were stimulated with 405nm light (single pass with $0.51 \mu$sec pixel dwell) to selectively activate defined regions within the peripheral ER. Cells were tracked for at least 2 minutes after stimulation with constant acquisition ($0.629$ sec/frame) to track diffusion of photoactivated signal into the surrounding ER network.


\subsection*{Image analysis and network structure extraction}

The machine learning segmentation toolkit ilastik~\cite{berg2019ilastik} is employed to segment ER network structures from live-cell images using the mcherry\_KDEL marker. A custom-written skeleton tracing subroutine in Matlab~\cite{MATLAB:2021} is used to extract a network structure from the probability file output by ilastik. This code is publicly provided at \url{https://github.com/lenafabr/networktools} and includes data structures for storing the network morphology as a set of nodes connected by edges with curved spatial paths. The networks are manually curated (using a network editing GUI provided as part of the networktools package) to remove unphysical terminal nodes arising from skeletonization artifacts.

For the calculations shown in Fig.~\ref{fig:MFPTs}A,C,D,F, and G a circular region of radius $20\mu$m is cut out from the extracted network structure. In Fig.~\ref{fig:MFPTs}B, eight sections of peripheral ER with radius $8.5\mu$m are used from three different cells. These eight extracted ER networks are again used in the pairwise reaction simulations (Fig.~\ref{fig:pairReactions}B-E) and in the Supplemental Material.


\subsection*{Analysis of photoactivated spreading data}


Imaging datasets for 9 individual cells are selected for analysis, each of which has a photoactivation region surrounded by a well-defined tubular network structure with primarily 3-way junctions.
The net signal over time is computed in 10 distinct wedge regions comprising an annulus around the photoactivation region with inner radius $3.5\mu$m and outer radius $6\mu$m. 
The photoactivated signal in wedge $j$ at time $i$ is defined as $w_{ij}^\text{PAGFP}$. The fractional signal is then given by $f_{ij}^\text{PAGFP} = w_{ij}^\text{PAGFP} / P_0^\text{PAGFP}$ where $P_0^\text{PAGFP}$ is the total initial signal within the photoactivated zone. 
 We find the slope (`signal arrival rate') of the fractional signal via a linear fit over the first $10$ seconds of imaging time following photoactivation.

Two rounds of filtering are applied to ensure a meaningful relationship between the photoactivated signal dynamics and the observed network structure. The first filter removes regions with extremely rapid and/or large fluctuations in the ER signal.
We calculate the time-variance in the fractional signal in each wedge as
\begin{equation}
V_{j} = \text{var}_i(w_{ij}^\text{PAGFP}/m_i^\text{PAGFP}),
\end{equation}
where $m_i^\text{PAGFP} = \sum_j w_{ij}^\text{PAGFP}$ is the total signal in the annular region at time $i$. Given the distribution of these time-variances, a threshold of $2.5 \times (\text{MAD} \times 1.4826)$, where MAD is the median absolute deviation, is used to define outliers with extreme ER dynamics, in keeping with commonly used outlier detection methods~\cite{leys2013detecting}.
%

The second round of filtering removes instances where network extraction does not accurately capture the underlying ER morphology. For example, small peripheral sheet regions, expanded junctions, or dense tubular matrices~\cite{nixon2016increased} can complicate the extraction of a well-defined network structure. Outlier regions are defined as those where the extracted total tubule length and the ER marker (mCherry\_KDEL) signal levels are mismatched. Specifically, we compute the time-averaged fractional mCherry\_KDEL signal for each wedge region in each cell as $s_j =  \left<w_{ij}^\text{mCherry}/m_i^\text{mCherry}\right>_i$. A linear fit is performed relating $s_j$ with the total extracted network length for each wedge, averaged over time. Any wedge with a residual above $2.5 \times (\text{MAD} \times 1.4826)$ is filtered out of the analysis.
Wedge regions filtered out due to either criterion are shown as gray dots in Fig.~\ref{fig:photoactivation}E-G.

\subsection*{Simulations of photoactivation on static networks}

For each frame in a photoactivation video, the ER network structure is extracted from the mCherry\_KDEL fluorescence channel, as described in the Image Analysis section. 
On each individual network structure, simulations of diffusing particles are conducted via a kinetic Monte Carlo method, as described in prior work~\cite{scott2021diffusive}. Briefly, analytically computed propagator functions are used to sample the time required for each particle to transition between neighboring nodes and edges, obviating any artifacts associated with a fixed time discretization. This method allows the particle to propagate in larger timesteps than would be achievable through classic Brownian dynamics simulations on a network.


Batches of $N = 10000$ particles are initiated within the experimentally photoactivated region, a $3 \times 3\mu\text{m}$ patch in the peripheral ER. Particles propagate through the network with a diffusivity of $D=1\mu\text{m}^2/\text{s}$, consistent with previous measurements of ER membrane protein diffusivity~\cite{sun2022unraveling}. All particle positions are saved at a frame rate matching the experimental imaging rate, $dt = 0.629\text{s}$.

To process the simulated data, we define individual wedge regions of the same size and location as in the experimental images and analyze the number of particles in each. Note that each simulation is run on a static network extracted from a single frame ($k$) of the experimental image. The simulated signal in each wedge ($w_{kij}^\text{sim}$) is then defined as the total number of particles in wedge $j$ and time point $i$ on the network extracted from frame $k$, and the fractional signal (plotted in Fig.~\ref{fig:photoactivation}D) is $f_{kij}^\text{sim} = w_{kij}^\text{sim}/N$. 

We next average this fractional signal over the different networks, defining $f_{ij}^\text{sim} = \left<f_{kij}^\text{sim}\right>_k$ (see Supplemental Material Fig.~S2A.iii for example averaged signal vs time curves). The resulting values are used to find the signal arrival rate (slope over first 10 seconds), exactly as for experimental data. Alternative methods for incorporating the time-varying ER network structure are considered in the Supplemental Material.

The simulations make it possible to incorporate a range of values for the particle diffusivity. We accomplish this by rescaling the simulation time in our analysis, which leads to a rescaling of the diffusivity (assuming static networks). For example, to test whether $D_\text{eff}=0.5\mu\text{m}^2/$s is a better representation of the protein diffusivity than $D_\text{orig}=1\mu\text{m}^2/$s, we can find the slope of $f_{ij}^\text{sim}$ over time $T_\text{scale} = \frac{D_\text{eff}}{D_\text{orig}}\times 10\text{s} = 5$s. The slope is then multiplied by $\frac{D_\text{orig}}{D_\text{eff}}$ to arrive at a simulated arrival rate with an effective diffusivity of $D_\text{eff}=0.5\mu\text{m}^2/$s.

We perform a linear fit of the rescaled simulated rates to the experimental protein arrival rates (slopes over $10$s). Repeating over a range of effective diffusivities, the value of $D_\text{eff}$ with the optimal fit indicates the best estimate of ER membrane protein diffusivity given the photoactivated spreading data.


\subsection*{Minimal model for dynamic ER networks}
To estimate the effects of ER network rearrangements on particle spreading, we conduct Brownian dynamics simulations on synthetic dynamic networks. 
To represent the dynamic network, we use a modified version of the previously published `minimal network model'~\cite{lin2014structure,lin2017modeling}. In this model, the network consists of mobile nodes connected by edges, where the node positions $\mathbf{x}_i(t)$ obey an overdamped Langevin equation
\begin{equation}
\label{eq:overdampedLangevin}
\frac{d \mathbf{x}_i }{dt} = -b \nabla f(\mathbf{x}_i) + \sqrt{2D_n}\boldsymbol{\eta}(t),
\end{equation}
where $D_n\approx 10^{-3}\mu\text{m}^2/$s~\cite{lin2017modeling} is the node diffusivity, $b$ is the node mobility in units of $\mu$m/s, and $f(\mathbf{x}_i)$ is the total edge length attached to each node. Specifically, $f(\mathbf{x}) = \sum_{j=1}^{d} \left| \mathbf{x} - \mathbf{y}_j \right|$, where the sum is over neighbor nodes and $\mathbf{y}_j$ are the neighbor positions. The stochastic variable $\boldsymbol{\eta}(t)$ is a Gaussian distributed noise term with mean zero and standard deviation $1$. 
 This model represents a network of edges which are under a constant tension, driving a minimization of their length.

As the edges of the network shrink, neighboring nodes approach each other. When two nodes are sufficiently close together, topological rearrangements of network connectivity can occur. If the two nodes are both degree 3 junctions, they undergo a T1 rearrangement~\cite{weaire2001physics} if and only if this decreases the total edge length. If one of the nodes has degree 2, or if they are connected by two edges (forming a short loop), then the two nodes can fuse together into a single node. The combination of these processes allows for ring-closure events, as observed in live-cell imaging of ER dynamics~\cite{lee1988dynamic,westrate2015form}.


To maintain a steady-state network structure, new edges are generated by a tube spawning and growth process. A new tube spawns at a fixed growth rate per existing total edge length ($k$, units of $\text{s}^{-1}\mu\text{m}^{-1}$). The new tube location is uniformly selected along existing edges. The nascent tube grows at a right angle from the parent edge, with fixed velocity $v=2\mu$m/s, comparable to rapid rates observed in dynamic ER images~\cite{lu2020structure,westrate2020vesicular}. 
When the tip of a nascent tube crosses an existing tube, it stops growing and fuses to form a new junction node.

Similar to \cite{lin2014structure}, the balance between new tubule growth and shrinking due to length minimization enables the dynamic network to reach a stable steady-state. Of the parameters in the model, the diffusivity is sufficiently low ($D \ll v \ell$, where $\ell \approx 1\mu$m is the characteristic edge length) to have little effect on network structure. Additionally, the tubule growth speed ($v\gg b$) is high enough that newly spawned tubules fuse much quicker than the node rearrangement timescales.
The network structure is thus largely determined by the remaining two parameters $k, b$, which are set to match observations of COS7 ER in live-cell images.

 The approximate growth rate $k$  is extracted from videos of COS7 peripheral ER, labeled with 0.2$\mu$g KDEL\_Venus (transfected as described above) and imaged on the Zeiss LSM 880 at a frame rate of $0.315$s, by manually counting new growth events (Supplemental Video 3). The number of growth events in a region of size $10\text{x}10\mu$m is manually counted over time interval 63s. This number is normalized by the time interval and the time-averaged total segmented ER length within the region, giving: $ k \approx 0.005 \mu\text{m}^{-1}\text{s}^{-1}$.

Dimensional analysis indicates that the average steady-state edge length in the network scales as $\ell \sim \sqrt{b/k}$. We tune the node mobility $b$ to set a typical ER network edge length $\ell \approx 1\mu$m~\cite{sun2022unraveling}, corresponding to an estimate of $b=0.05\mu\text{m}/\text{s}$.

The resulting dynamic network model thus has tubule lengths and turnover timescales that approximately represent those of the COS7 ER (`normal ER model'). For comparison, we consider also a model where $b$ and $k$ are both increased by $2\times$, allowing for more rapid turnover but the same steady-state structure (`fast ER model').


\subsection*{Simulating photoactivated spread on the dynamic network model}
Diffusive particles ($N = 10000$) are simulated on the dynamic network using Brownian dynamics, with particles moving along the network edges in discrete timesteps $dt=10^{-3}$s, with diffusivity $D=1\mu\text{m}^2/$s, and network structure updated at each timestep. After each network update, the particle position is projected to the closest location in the new network. The network architecture is first evolved for a total time of 1000s to allow it to reach steady state prior to initiating the diffusive particles. The particles are placed within a square 3x3$\mu$m region of the network and the joint simulations of particle and network evolution then proceed for an additional $15$s of simulated time.

The number of particles arriving in each wedge region surrounding the starting center is analyzed on both the dynamically evolving network and on each individual static network structure extracted at $0.6$s intervals from the simulation. The signal arrival rates are obtained as described in the previous section.

%
%


\subsection*{Paired particle simulations}

Simulations of reactive particle pairs are run using the propagator-based approach, which enables particles to hop rapidly from node to node of the network until they come within the same neighborhood of each other. Details of the methodology, including the appropriate propagator functions for two reactive particles on the same edge, are provided in prior work~\cite{scott2021diffusive}. Each simulation is run until the two particles encounter each other, and the reaction position on the network is recorded. A total of $N=160000$ particles are simulated on each network structure. The ER network structures are extracted from COS7 cell imaging data as described in the Image Analysis section.

Two other families of network are also analyzed. Eight circular honeycomb networks are generated, each with the same diameter and total edge length as one of the eight ER structures analyzed. 
Mikado networks are generated by scattering $N_{\text{rod}}$ randomly oriented rods of length $L_\text{rod}$ in a square of size $L_\text{space}\text{ x }L_\text{space}$. The intersections of these rods define the nodes of the network, the segments of rods between intersections define the edges of the network. This algorithm generates highly heterogeneous networks with a density that is tunable by changing any of the three input parameters~\needcite. However, Mikado networks tend to have degree 4 junctions, whereas ER (and honeycomb) networks are composed of mostly degree 3 junctions. Our modification to the Mikado networks is thus to remove degree 4 nodes by iteratively removing 1 random edge from a randomly chosen degree 4 node until all nodes in the network have degree 3 or less.
The Mikado parameters are chosen to be $N_{\text{rod}}=80$, $L_\text{rod}=12\mu$m, and $L_\text{space}=24\mu$m and a circular portion of the network is extracted with diameter $18\mu$m, matching the circular ER networks. We generate many copies of these circular modified Mikado networks and select for analysis only those which have a total length within 
 $5\%$ of the corresponding ER network.

\subsection*{ERES localization on ER network}
COS7 cells were seeded in plastic 6-well dishes and transfected as described in the photoactivation experiment section. Cells were then imaged as previously described~\cite{westrate2020vesicular}. The following standard amounts of DNA were transfected per mL: 0.1$\mu$g GFP\_Sec16s, 0.1$\mu$g GFP\_Sec23a, 0.1$\mu$g GFP\_Sec24d, 0.1$\mu$g EYFP\_Sec31a, 0.2$\mu$g mcherry\_KDEL. Images were acquired on inverted fluorescent microscope (TE2000-U; Nikon) equipped with a 100x oil objective (NA 1.4) on an electron-multiplying charge coupled device (CCD) camera (Andor). Live cell imaging was performed at $37^\circ$C following media change to pre-warmed imaging media (fluorobrite DMEM (Invitrogen) + 10\% FBS).

Images of $23$ different COS7 cells are analyzed as described in the Image Analysis section to extract the peripheral ER network structure. The perinuclear region is manually excised from each one. ERES locations are identified as puncta in the GFP or EYFP fluorescent signal using a previously published implementation of the standard particle localization algorithm by Crocker and Grier~\cite{crocker1996methods,koslover2016disentangling}. The ERES positions are projected onto the nearest point along the extracted network structure. The global mean first passage time (GMFPT) to each projected ERES location is computed analytically as previously described~\cite{scott2021diffusive}. A total of 1466 exit site positions are analyzed. For comparison, $1000$ target points are selected uniformly at random along the edges of each network structure, and the GMFPT is computed to each of those points individually.

%

\section*{Results and Discussion}
\label{sec:results}

\subsection*{ER network structures exhibit spatially heterogeneous accessibility}
\label{sec:analytics}

\begin{figure*}[hbt!]
	\centering
	\includegraphics[width=0.95\textwidth]{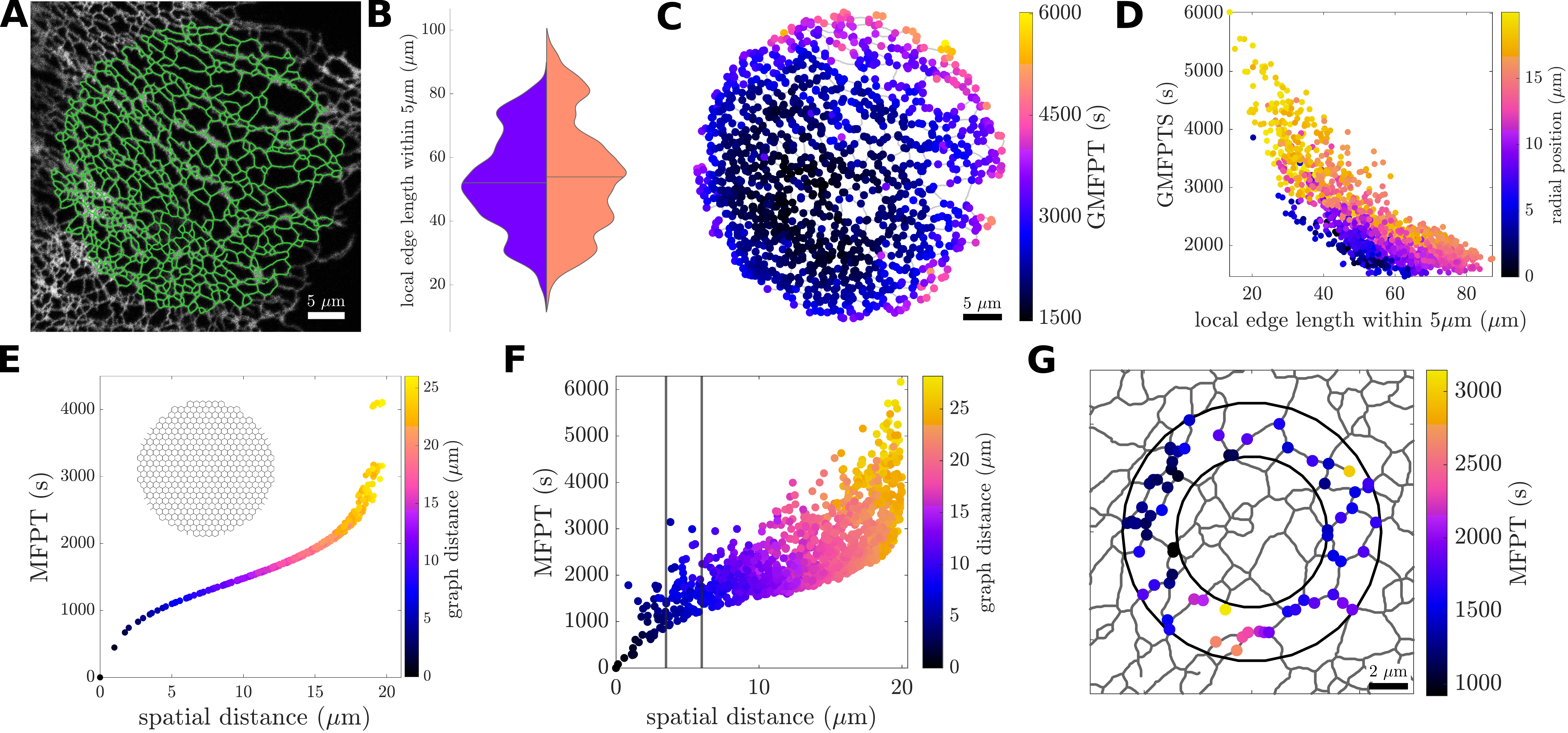}
	\caption{ (A) Confocal image of COS7 cell expressing fluorescent ER marker (KDEL\_Venus, gray) with extracted network structure overlay (green). (B) The distribution of local edge lengths in one cell (left) is similar to the corresponding distribution across multiple cells (right). (C) Global mean first passage times (GMFPTs) to nodes on the example COS7 ER network shown in A. (D) GMFPT scales inversely with local edge length, color denotes radial position from center of network. (E) MFPT to each network node for particles diffusing outward from the center of a circular honeycomb network. (F) MFPTs to all network nodes for particles diffusing outward from the center of the example ER network. Vertical lines highlight heterogeneity in a ring from $3.5-6\mu$m around the center. (G) MFPTs for nodes  in the ring from $3.5-6\mu$m, as highlighted in F.
 	}
	\label{fig:MFPTs}
\end{figure*}

The peripheral endoplasmic reticulum forms an intricate web of tubules, with primarily three-way junctions scattered at varying densities across the cell periphery. We aim to characterize the heterogeneity of the network structure and its effects on the accessibility of different regions by particles diffusing on the network.

ER network morphologies are extracted from confocal images of the peripheral ER in cultured COS7 cells (Fig.~\ref{fig:MFPTs}A), where these network structures are largely planar. The network structure is simplified into effectively one-dimensional edges (not necessarily straight), connecting point-like nodes. Although more complex peripheral structures, including hole-studded sheets~\cite{schroeder2019dynamic} and dense localized matrices\cite{nixon2016increased}, have been observed, we focus here on regions composed primarily of well-defined tubules and junctions.

The ER density in different spatial regions can be characterized by computing the local edge length $L_\text{loc}(x; \sigma)$, defined as the total length of network tubules that falls within distance $\sigma = 5\mu$m of position $x$. We sample local edge length for random points scattered across the domain of an example network (shown in Fig.~\ref{fig:MFPTs}A). The values of $L_\text{loc}(x; \sigma)$ span one order of magnitude, demonstrating substantial spatial heterogeneity in the ER density (Fig.~\ref{fig:MFPTs}B). Notably, spatial variation in the local edge length within a single network is comparable to the variation between networks extracted from different cells.


While local edge length provides a purely structural metric of heterogeneity, we further consider the consequences of network variability on the diffusive transport of particles within the ER. 
One useful metric for quantifying search efficiency on spatial networks is the global mean first passage time (GMFPT)~\cite{brown2020impact}, which gives the mean first passage time for a diffusing particle to reach a given node in the network, averaged over all starting nodes. This quantity can be computed analytically from the edge lengths and topology of the network~\cite{scott2021diffusive}.

The GMFPTs for different nodes in a single ER network can vary substantially (Fig.~\ref{fig:MFPTs}C). Nodes near the boundary have a higher GMFPT, whereas more centrally located nodes and those in denser regions of the network exhibit the lowest GMFPTs. Some of the variation in GMFPT can be explained by the local edge length surrounding a node, as well as proximity to the edge of the domain (Fig.~\ref{fig:MFPTs}D), both a measure of centrality within the network~\cite{barthelemy2011spatial}. However, even nodes with similar local edge lengths and radial position can have GMFPTs that vary by a factor of 2.
We note that the ER networks form a highly-looped structure composed primarily of 3-way junctions, with less than 5\% terminal nodes. Thus, although network search times are known to vary with node degree~\cite{weng2017hunting}, the degree of each node is insufficient to account for the observed variability of the GMFPTs. 


Individual mean first passage times (MFPTs) between pairs of nodes in the network can be used to further assess heterogeneity in local transport processes.
The MFPT for a particle diffusing outward from a central point to each possible target node in a uniform honeycomb network exhibits a characteristic scaling with distance, as shown in Fig.~\ref{fig:MFPTs}E. 
Unsurprisingly, nodes that are located farther from the source tend to have higher MFPTs, with the search time increasing exponentially for the most distant population of nodes.
This particular scaling of the diffusive search time relative to distance has also been observed for particles that hop actively across edges in planar network structures~\cite{dora2020active}.
Similar scaling is found in diffusive search for targets on the ER (Fig.~\ref{fig:MFPTs}F).
However, the heterogeneity of the ER network structure gives rise to a broad range of mean search times for nodes at similar distances from the source. A factor of 3 difference is observed for nodes that fall within a ring from $3.5\mu\text{m}-6\mu\text{m}$ from the center (Fig.~\ref{fig:MFPTs}F,~G).





Overall, we use analytic mean first passage times as a measure of accessibility for different regions of the ER, either by particles starting throughout the network, or those originating from a localized source. This accessibility is shown to vary between different regions of an ER network, due to the heterogeneous density and connectivity patterns of the tubules. 

\subsection*{Network morphology governs the nonuniform spread of photoactivated proteins}
\label{sec:paspread}

\begin{figure*}[hbt!]
	\centering
	\includegraphics[width=0.95\textwidth]{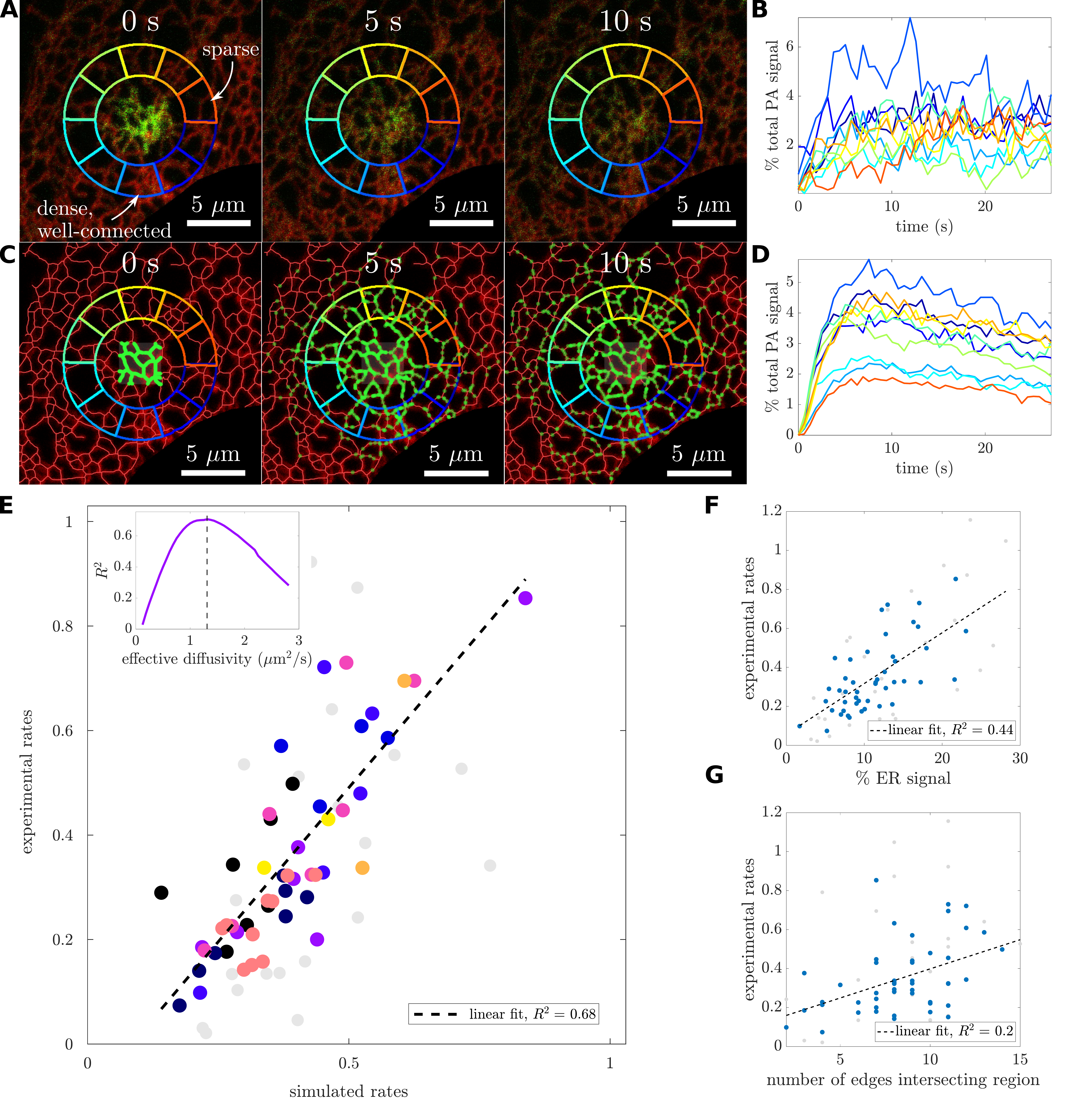}
	\caption{Spreading of localized bolus of particles over the ER network. (A) ER membrane protein PAGFP\_Calnexin (green) is pulse-activated in a local region, while ER luminal marker mCherry\_KDEL (red) serves to visualize network structure. Equidistant surrounding regions (colored wedges) are used to analyze signal spread. (B) Photoactivated signal arriving in each analyzed region, normalized by initial total signal in photoactivated zone. (C) Snapshots of simulations on frozen ER structures extracted from first frame in A. (D) Simulated particle counts arriving in individual analysis regions, normalized by total number of particles. (E) Correlation between signal arrival rates (slopes of signal vs time curves) for experimental and simulated data. Color indicates cell (N=9). Inset: simulated protein arrival rates best match experimental arrival rates when effective diffusivity is scaled from $D_\text{orig}=1\mu\text{m}^2/$s to $D_\text{eff}=1.3\mu\text{m}^2/$s (dashed line). (F) Correlation of experimental signal arrival rate in individual regions versus the fraction of ER marker signal in that region. (G) Correlation of experimental signal arrival rate with the number of edges intersecting the boundary of each region. Regions removed due to filtering are shown in gray in E-G.}
	\label{fig:photoactivation}
\end{figure*}

Although mean first passage times are a convenient, easily computed metric of diffusive accessibility, they are difficult to probe experimentally. To directly observe the heterogeneity of diffusive spreading within the ER, we consider instead the short-time rate of arrival to nearby regions surrounding a particle source. This process is visualized by photoactivating ER membrane-associated proteins within a localized region of the network and watching their spread into surrounding regions.



Cultured COS7 cells are transfected with PAGFP\_Calnexin, a membrane-bound ER protein in the ER with a photoactivatable fluorescent tag, as well as mCherry\_KDEL as a general marker for ER structure. A single pass photoactivating pulse is applied in a $3\mu\text{m} \times 3\mu\text{m}$ square of the peripheral ER.
 Several frames from an example video (Supplemental Video 1) are shown in~Fig.~\ref{fig:photoactivation}A with mCherry\_KDEL in red and the PAGFP\_Calnexin in green. The initial dense bolus of photoactivated proteins can be seen spreading outward through the network. We track the signal in individual small regions located equidistant from the photoactivation site. Diffusive spreading of particles over a homogeneous continuum would be expected to yield similar time-courses of signal arrival to each of these regions. However, the observed PAGFP fluorescence signal over time varies substantially between the individual wedges in a single cell (Fig.~\ref{fig:photoactivation}B).
 This variability can be attributed to the heterogeneous distribution and connectivity of the ER tubules.
  Intuitively, the blue region contains dense, highly connected tubules and has the strongest and fastest-growing photoactivation signal. By contrast, the orange region is poorly connected to the activation site and exhibits the smallest initial signal growth.

To account for the observed differences in signal arrival rates due to ER morphology, we extracted the ER network structure in the vicinity of the photoactivation site and carried out agent-based simulations of diffusing particles initiated at the site (Fig.~\ref{fig:photoactivation}C, Supplemental Video 2). Quantifying the number of simulated particles accumulating in each region over time allows for a direct comparison between simulated and observed fluorescent signal. For both the experimental and simulated data, we normalize the measured signal in each region by the initial total signal within a disc of $3.5\mu$m radius centered on the photoactivation zone (inner circle in Fig.~\ref{fig:photoactivation}A,C). Thus, the reported signal traces are given in terms of the fraction of initially photoactivated particles present in a given region at a given time.
%
The normalized simulated signal (Fig.~\ref{fig:photoactivation}C) exhibits similar behavior to the experimental results, with well-connected dense regions receiving more signal faster than poorly-connected and sparse ER regions.

To partially incorporate the effect of ER network rearrangement over time, the photoactivation simulations are run on network structures extracted for every frame of the experimental movie (at time interval $0.6$s). The signal over time is then averaged across the ensemble of simulations on all of these different network structures.
This ensemble-averaged simulated signal is used in the subsequent analysis. Analogous results using only a single network structure can be found in the Supplemental Material.

To quantitatively compare protein arrival rates in the experimental and simulated ER networks, we extract the slope of the normalized signal curves up to $10$ seconds following photoactivation. These slopes (referred to as `arrival rates') serve as a simple metric that provides information about the spatial heterogeneity of protein spreading around the photoactivation site. 
%
%
%
Because our simulations are carried out on network structures extracted from the experimental images, it is possible to directly compare the rate of signal arrival in matched regions between experimental and simulated data (Fig.~\ref{fig:photoactivation}E). Regions where the extracted network length was a poor match for the observed ER marker (mCherry\_KDEL) fluorescence, or where the ER marker showed large fluctuations over time, were filtered out of the analysis (gray dots; see Methods for details).
Notably, the variability of measured rates between regions within each individual cell (same color dots) is comparable to the inter-cell variability (different color dots), indicating that the arrival rates are similarly heterogeneous in all the observed cells.
The experimental and simulated arrival rates show a direct correlation: $R^2 = 0.68$, obtained from a linear fit. The high correlation implies that diffusive particle motion over an ER network is a good predictor of signal arrival to different regions.

Notably, the simulation time can be rescaled to effectively represent particles of different diffusivity (see Methods for details). We compare the correlation of signal arrival rates between experimental measurements and simulations with different time-scaling. The simulations which best correlate with experimental values correspond to a particle diffusivity of $D_\text{eff} \approx 1.3\mu\text{m}^2/\text{s}$ (Fig.~\ref{fig:photoactivation}E, inset), a value that is similar in magnitude to previous measurements of diffusivity via single-particle tracking for other ER membrane proteins~\cite{sun2022unraveling}. This result demonstrates that particle diffusivity in the ER can be measured by quantifying signal arrival rates to different structural regions regions of network, all located relatively close to the photoactivated zone, without the need for tracking longer-range spread across the cell~\cite{konno2021endoplasmic}. 

In order to test whether particle diffusion simulations are more predictive than simpler metrics of network structure, we also compare the experimental arrival rates to the mCherry\_KDEL ER signal in each region. A linear fit (Fig.~\ref{fig:photoactivation}F) demonstrates there is some correlation between the two ($R^2=0.44$), but variation in the ER volume within each region (as measured by mCherry\_KDEL signal) cannot capture the full variability in protein spreading rates. A simple metric for local connectivity, the number of edges crossing the boundary of each wedge region, is shown to be roughly correlated (Fig.~\ref{fig:photoactivation}G, $R^2=0.2$), but also does not provide a strong predictor of protein arrival rates. Thus, the distribution of protein spreading rates in live cells is best modeled by simulations which take into account not just the local ER density in a region, but also the connectivity of the surrounding network together with the dynamics of diffusive particles moving through this network.


\begin{figure*}[hbt!]
	\centering
	\includegraphics[width=0.75\textwidth]{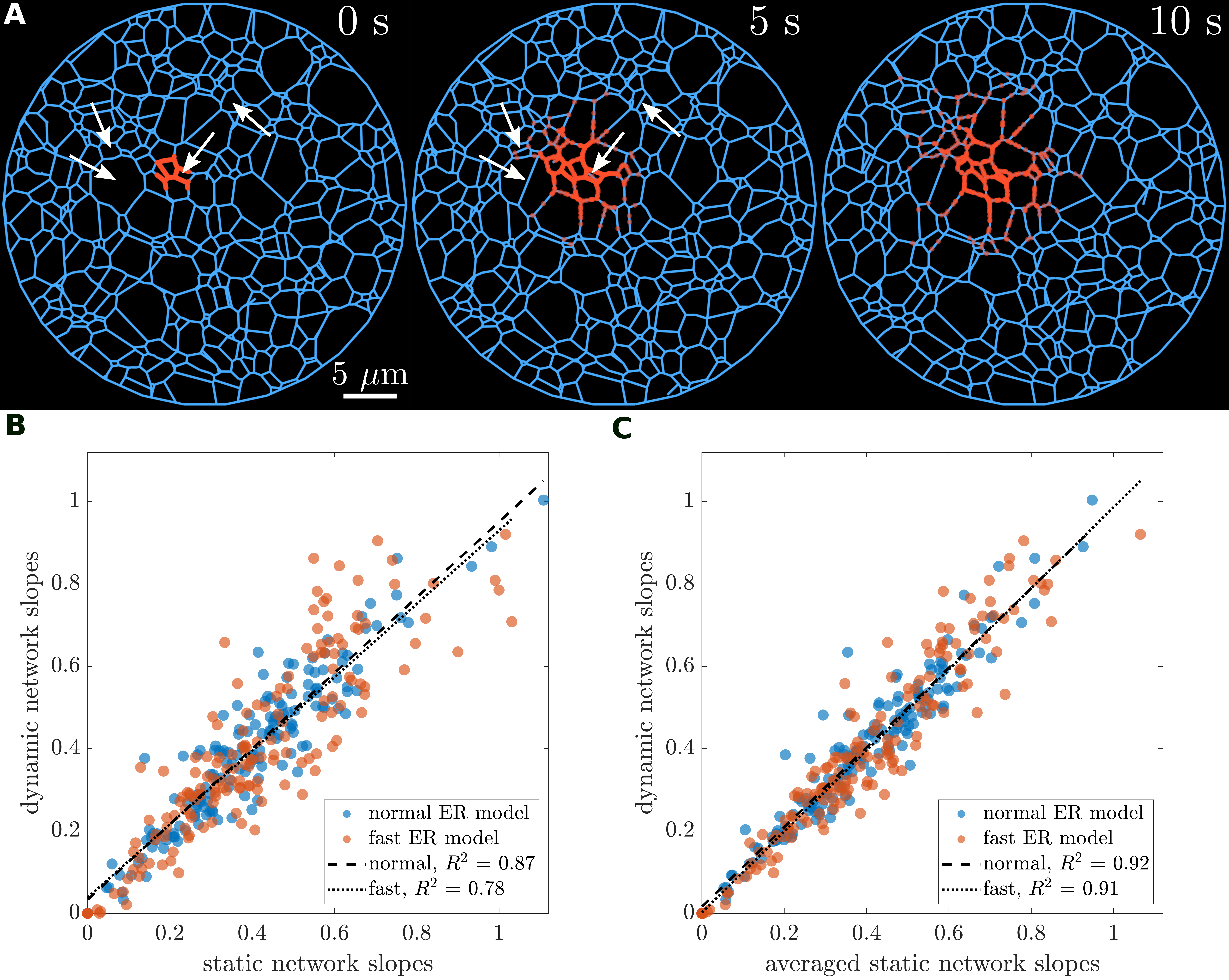}
	\caption{ER network dynamics does not substantially affect particle spreading. (A) Snapshots from simulation of diffusing particles spreading from a local region, on a minimal network dynamic model. Network edges are blue and a subset (500) of the simulated particles ($D=1\mu\text{m}^2/$s) are shown in orange. White arrows highlight several new edges that grew between first and second snapshots. (B) Correlation of signal arrival rate (slope of signal-vs-time curves) to individual regions, comparing simulations on a single static network structure and on dynamic minimal network model with turnover timescales comparable to ER dynamics (blue) or $2\times$ faster (red). (C) Comparison of signal arrival rates for simulation on a dynamic network versus simulations of particles diffusing on a static network, averaged over static structures from individual snapshots of the minimal network. The average static rates are obtained in the same manner as in the analysis of experimental data.}
	\label{fig:corrplots}
\end{figure*}

\subsection*{Slow ER network dynamics have little effect on particle spreading}

The ER network in a living cell is itself a dynamic structure, with network rearrangements occurring over tens-of-second timescales as a result of attachment to motile organelles, molecular motors, and growing microtubule tips~\cite{westrate2015form,friedman2010er,lu2020structure}. In comparing the measured rates of protein spread to simulations of diffusing particles (Fig.~\ref{fig:photoactivation}E), we account for time variation in ER architecture by averaging over network structures extracted from each frame.

To gain a better sense of how ER tubule dynamics may contribute to the  spread of photoactivated proteins, we incorporate network rearrangements directly into our simulations, by treating the ER as a `minimal network' with tubules subject to growth and constant tension~\cite{lin2014structure}. These synthetic dynamic networks (described in the Methods) mimic the rearrangements of the ER over time, including new tubule growth, junction sliding, and the merging of junctions.






The two parameters primarily responsible for determining the equilibrium properties are node mobility (units of $\mu$m/s, sets speed with which nodes rearrange) and new tubule growth rate (units of $\mu\text{m}^{-1}\text{s}^{-1}$, rate at which new tubules are pulled out of existing tubules). Other parameters, such as node diffusivity and new tubule growth speed play a secondary role in the parameter regimes considered here.
Input parameters to the model are set so that network properties at equilibrium match the ER network in COS7 cells. Specifically, the modeled networks match the measured rate of new tubule formation (Supplemental Video 3) and the steady-state average edge length in the network (see Methods for details). For comparison, we also ran simulations of networks that exhibit faster dynamics, with both tubule growth rate and node mobility increased by a factor of $2$. These faster networks have the same steady-state network structure but rearrange twice as rapidly.

For each of these dynamic networks, 16 separate photoactivation events are simulated in different regions of the network. Particles are initiated within $3\times3\mu \text{m}$ patches and allowed to diffuse through the structure either on a static network or concurrently with the network dynamics (Fig.~\ref{fig:corrplots}A, Supplemental Video 4). We compare the rate of particles arriving to equidistant regions surrounding the initiation zone both with and without network dynamics.

When the particle simulations are run on a single static network structure, the arrival rates are moderately well-correlated ($R^2 = 0.87$) with the rates observed for simulations on dynamic networks (Fig.~\ref{fig:corrplots}B). Faster dynamics in the synthetic networks reduces this correlation to $R^2 = 0.78$. Intuitively, as the rearrangements occur more quickly, diffusive particles encounter more extensive changes in structure during the 10-second timescale of the measurement. 

Notably, even when network dynamics are twice as rapid as the experimentally observed dynamics of the ER network, the static network approximation is a good predictor for particle arrival rates. We can estimate the importance of active network rearrangements versus the diffusive motion of the particles by considering an effective P\'eclet number for the system. The mobility parameter for the dynamic networks ($b = 0.05\mu$m/s) sets a typical velocity for tension-driven sliding. Over a length scale of $10\mu$m 
(corresponding to the diameter of the analyzed region), the corresponding P\'eclet number for a protein within the network is $\text{Pe} = vL/D \approx 0.5$. Doubling the rate of ER rearrangement doubles this P\'eclet number. Because this dimensionless quantity is close to or below $\text{Pe} = 1$, the motion of the particles is dominated by their diffusivity rather than by the tubule rearrangement dynamics.

The moderate effect of network dynamics on particle spreading can be partly accounted for by running simulations on many individual static network structures extracted at different points in time. We perform this analysis using snapshots of our simulated dynamic networks and averaging the signal in each region at each time point. For this ensemble-averaged data, the arrival rates on static and dynamic networks become more closely correlated (Fig.~\ref{fig:corrplots}C), even in the case of rapid network rearrangements ($R^2 =0.91$). Thus, the effect of network dynamics is almost entirely accounted for by averaging multiple static simulations on consecutive network structures. These results on synthetic dynamic networks validate the use of the same ensemble averaging approach when analyzing experimental data.

\subsection*{ER structure directs reaction locations}
\label{sec:reactions}

\begin{figure*}[hbt!]
	\centering
	\includegraphics[width=0.9\textwidth]{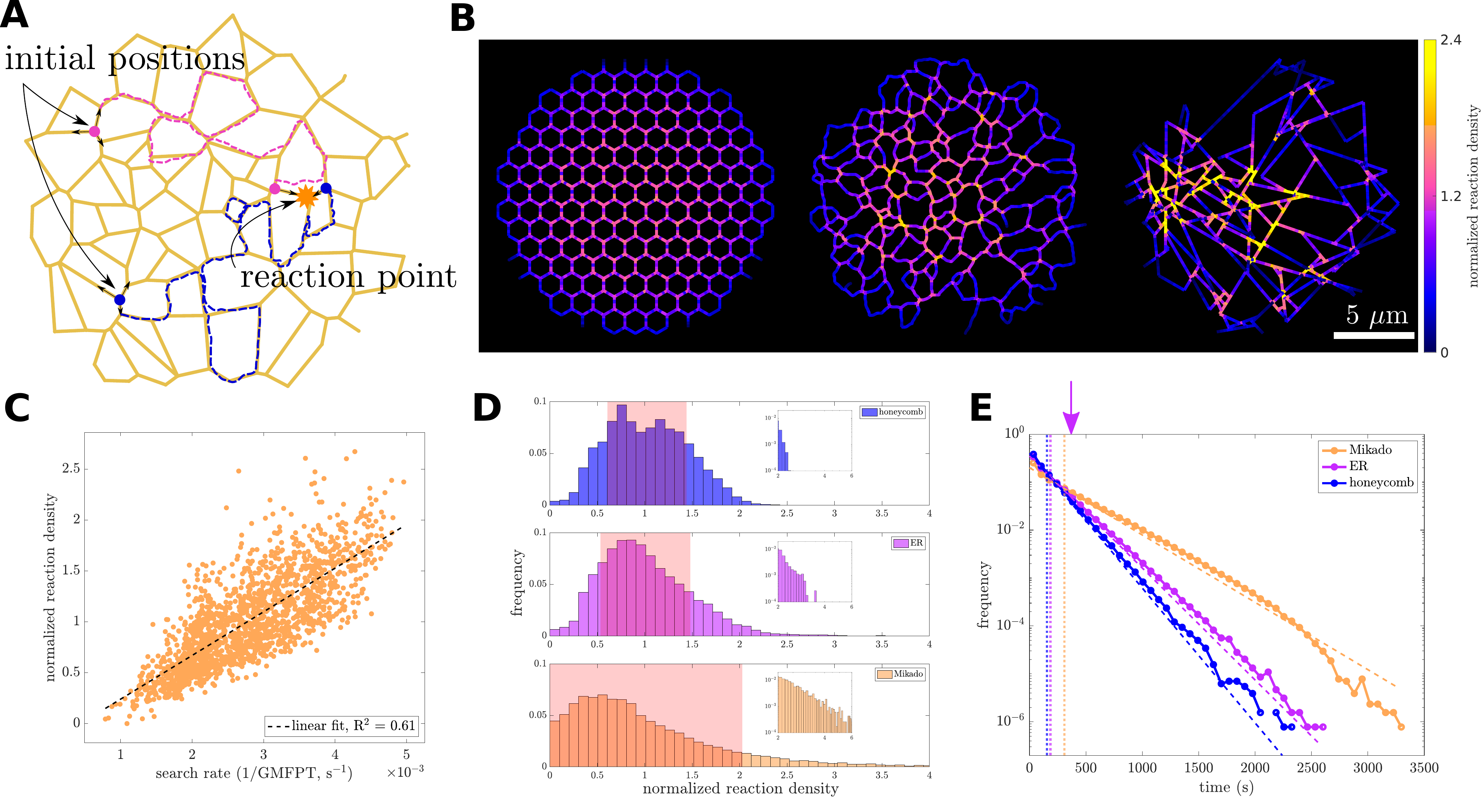}
	\caption{ER heterogeneity leads to hot spots of paired particle encounters. (A) Schematic of paired particle simulations. Pairs of particles (pink and blue circles) diffuse through the network (dashed lines indicate trajectories) until they encounter and react. (B) Normalized reaction density on three example networks, each with similar total edge length and spatial size. For each discretized segment of network, the fraction of simulated reactions occurring within that segment is normalized by the fraction of total edge length contained within that segment. Left panel, a homogeneous honeycomb network with the same average edge length as the ER network in the middle panel. Middle panel, ER network is extracted from a section of COS7 peripheral ER, and exhibits regions of higher reaction density than the homogeneous honeycomb (bright yellow segments). In the right panel, the normalized reaction density on a highly heterogeneous synthetic, Mikado-like network, exhibiting even more pronounced hot spots than the ER. (C) Paired reaction density on each segment of the ER is roughly correlated with the inverse of the global mean first passage time (GMFPT) to that segment. (D) Distribution of reaction rate densities for all discretized segments in honeycomb, ER, and Mikado-like networks, showing increasing heterogeneity in the densities. Insets show long tail of distribution plotted on log-log axes. Mean for each distribution is one, red overlay denotes standard deviation. (E) Distribution of paired reaction times in the three network structures. Dashed lines show fit to an exponential distribution. Dotted lines mark the mean reaction time. The target-averaged GMFPT on the ER networks indicated by the purple arrow is more than twice the mean pair reaction time.}
	\label{fig:pairReactions}
\end{figure*}

The endoplasmic reticulum does more than simply serve as a transport hub for proteins, lipids and ions; it also plays a role in protein synthesis and quality control~\cite{ellgaard2003quality,sontag2014sorting}, as well as forming functionally important contact sites with other organelles~\cite{wu2018here}. The formation of reactive complexes, exit sites for protein export, and contact site assemblies requires multiple intra-ER particles to find each other within the network.
In order to better understand how diffusion-mediated biochemical reactions are impacted by ER morphology, we simulate reactive particle pairs diffusing on extracted ER network structures (Fig.~\ref{fig:pairReactions}A). From these simulations, both the spatial locations of reactions on the network as well as the distribution of reaction times are extracted.


Many previous studies of diffusive processes on networks have focused on the temporal properties of reactions or exit times (e.g.\ MFPTs, extreme statistics, and full FPT distributions~\cite{scott2021diffusive,viana2020mitochondrial,lanoiselee2018diffusion}), without considering in detail where those reactions occur.
Here, we provide fresh insight by analyzing the spatial locations, as well as temporal distributions, of pair-wise reactions on the ER.
%
%
Pairs of particles are distributed randomly across the network to begin the simulation. Each pair diffuses along the edges of the network until the two particles come into contact with one another. At this point, they react and the position and time of reaction is recorded. The network edges are meshed into segments of length $\ell \approx 0.2\mu\text{m}$, and the normalized reaction density in mesh cell $i$ is defined as:
\begin{equation}
\begin{split}
	\gamma_i = \frac{\text{\# of reactions in cell $i$}}{\text{\# of particle pairs}}\times \frac{\text{total network length}}{\ell}.
\end{split}
\end{equation} 
When averaged over an entire network, $\left<\gamma\right> = 1$.
Simulations on the ER network structure demonstrate that paired particle reaction locations are heterogeneous (Fig.~\ref{fig:pairReactions}B, middle panel), with some regions showing a particularly high reaction density $\gamma$. Certain tubule segments are more likely to serve as the reaction site, due to their enhanced connectivity to the rest of the network. The normalized reaction density correlates with the inverse of the GMFPT (Fig.~\ref{fig:pairReactions}C), indicating that these highly reactive regions are in fact easier to find by diffusing particles.

These simulations imply that heterogeneity in ER structure and accessibility is expected to result in diffusive particle reactions becoming concentrated within certain regions. For comparison, we repeat the simulations on two synthetic network structures: a homogeneous honeycomb network (Fig.~\ref{fig:pairReactions}B, left panel) and a highly heterogeneous modified Mikado network~\cite{wilhelm2003elasticity} (Fig.~\ref{fig:pairReactions}B, right panel; see Methods for details). Both of these networks have the same spatial extent and total network length as the extracted ER networks. This allows for a quantitative comparison of the reaction density and reaction time distributions between all three families of networks.


As expected, reaction locations are more uniformly distributed on the honeycomb network. Within this homogeneous network, reactions are slightly more likely to occur at junction nodes rather than along the edges, in keeping with past work showing random walkers are more likely to encounter each other at higher-degree network nodes~\cite{weng2017hunting}. There is also a dearth of reactions at the network boundary, mirroring the increased GMFPT in the boundary region (Fig.~\ref{fig:MFPTs}C).
The ER networks show a similar drop-off in reaction density along edges as compared to junctions, as well as at the boundary. Moreover, due to the heterogeneous network density and connectivity, reactions are more concentrated into certain junctions within the network, with a higher maximum reaction density at these select junctions than is observed in the more uniform honeycomb. Reactions are further concentrated in the modified Mikado networks, demonstrating that more heterogeneous networks exhibit a broader range of reaction densities. This effect is quantified in Fig.~\ref{fig:pairReactions}D, where a longer tail is visible in the distribution of normalized reaction densities for ER and Mikado networks, as compared to the honeycomb. The morphology and connectivity of a network can thus tune the spatial distribution of reaction locations.


Network structure is not only responsible for shaping the spatial profile of reaction density, but can also affect the overall reaction time~\cite{brown2020impact,viana2020mitochondrial}.
The distribution of pair-wise reaction times on each network exhibits exponential scaling (Fig.~\ref{fig:pairReactions}E), as for a Poisson process with a single dominant time-scale. As noted in previous work, the mean reaction time on the ER (dashed purple line) is less than half of the target-averaged GMFPT (purple arrow)~\cite{scott2021diffusive}.
Even though there are higher spatial reaction densities in the more heterogeneous networks, mean reaction time is lowest in the homogeneous honeycombs and highest in the modified Mikado networks (Fig.~\ref{fig:pairReactions}E). Thus, there is a trade-off between locally concentrating reactions in space versus minimizing overall reaction time.



\begin{figure*}[hbt!]
	\centering
	\includegraphics[width=0.5\textwidth]{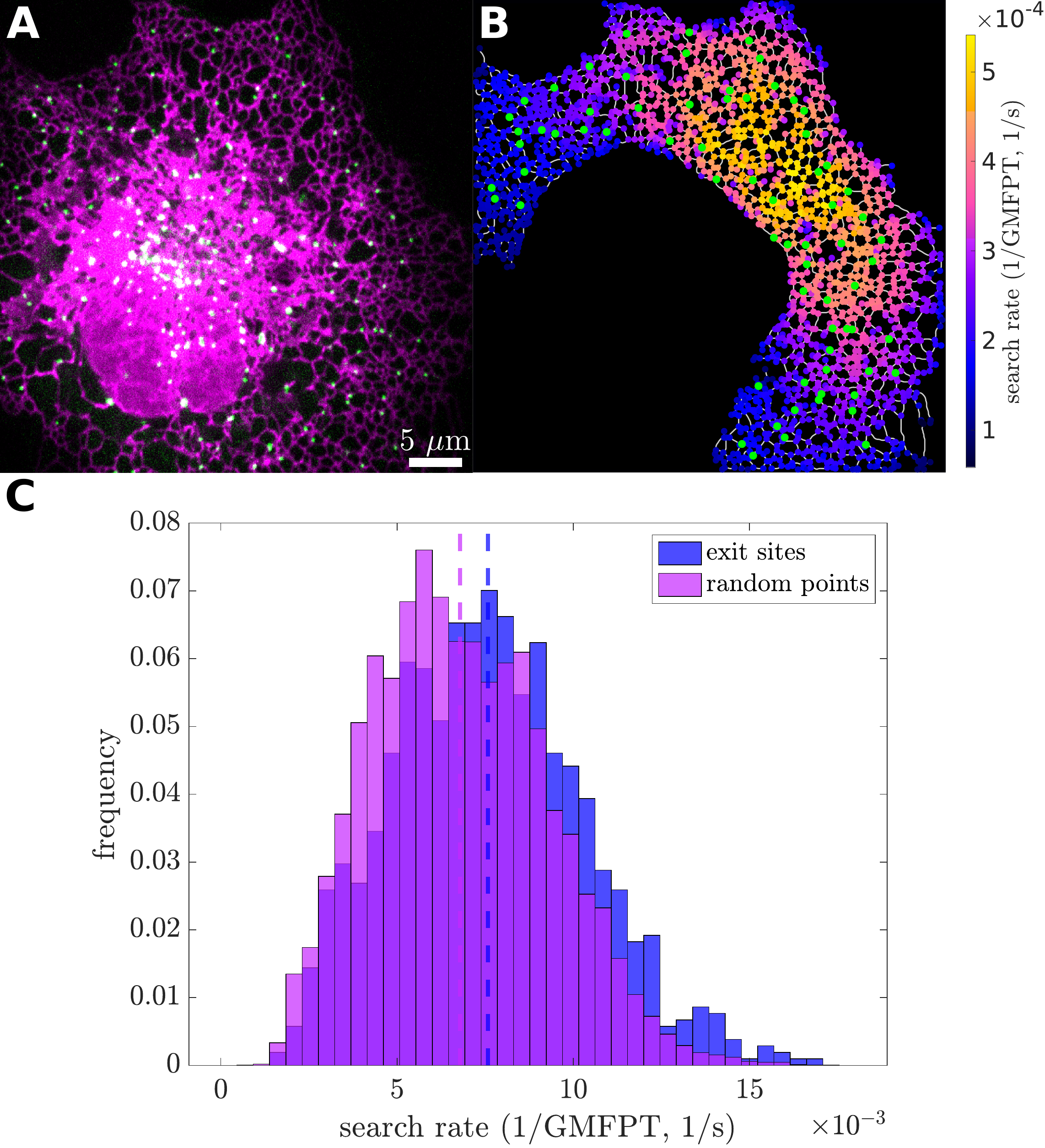}
	\caption{Concentration of ERES in regions of the peripheral ER with high search rate. (A) Image of the ER (magenta, mCherry\_KDEL) and ERES (green, GFP\_Sec24d) of a COS7 cell. (B) Extracted network structure of peripheral ER (excluding nucleus and perinuclear sheet regions). Junctions are colored by their effective search rate (inverse GMFPT, units of $1/$s). ERES positions are shown in green. (C) ERES positions exhibit higher effective search rates than random points on the network. Distributions consist of $1466$ exit sites and $23$k random points extracted from the peripheral ER of 23 different COS7 cells.}
	\label{fig:eresReactions}
\end{figure*}

Given that our simulation results demonstrate the presence of disproportionately reactive regions within the peripheral ER network, we sought to examine whether certain ER-associated protein assemblies may be more likely to localize to such regions. Specifically, we explore the distribution of ER exit sites (ERES), which serve as the export hubs for newly synthesized proteins in the ER~\cite{watson2006sec16,westrate2020vesicular}. The mechanism underlying the distribution of exit sites on the network is not well understood, although prior work has suggested they may arise from a process of confined diffusive aggregation~\cite{speckner2021unscrambling}. The process of ERES formation is not modeled explicitly here. Instead we investigate whether these structures are more likely to be found within highly connected and reactive network regions (as measured by search rate, defined as 1/GMFPT).

We extract ERES puncta locations using several different markers for the exit sites (Fig.~\ref{fig:eresReactions}A, details in Methods) and project these locations onto the extracted ER network structure (Fig.~\ref{fig:eresReactions}B). We next calculate the search rate to each ERES position, and compare the distribution of these rates to that expected for randomly selected locations along the networks (Fig.~\ref{fig:pairReactions}I).
The distribution of search rates at the exit sites (mean $\pm$ std: $6.8 \pm 2.8 \times 10^{-3}$s$^{-1}$) is shifted to higher values as compared to the randomized control (mean $\pm$ std: $6.1 \pm 2.7 \times 10^{-3}$s$^{-1}$). 
Given the large numbers of exit sites and random points sampled (1466 exit sites, 23k random points), this difference is statistically significant, $p<<10^{-6}$ by a one-sided student's t-test. This indicates that the ERES are disproportionately likely to be found in highly connected regions of the ER network.

These results indicate a potential structure-function relationship for the peripheral ER network. Structural heterogeneity in the network translates to heterogeneous locations for reactions of diffusing particles. In turn, certain multi-protein assemblies within the ER network appear to be localized to the more highly reactive regions, where they can be more easily reached by other diffusive particles.


%
%
%
%


\section*{Conclusion}


In this work, we highlight the heterogeneous connectivity of the tubular ER network and its consequences for diffusive particle transport. We extract peripheral ER network structures from live-cell confocal images of COS7 cells and analytically compute mean first passage times (MFPTs) for particles diffusing over these networks. These calculations allow us to quantify the variability in diffusive accessibility within individual ER architectures. The global MFPT to individual nodes within the network is found to vary by up to 4-fold due to the heterogeneous connectivity of the network.

We then directly visualize the local spreading of ER membrane proteins from an initial region of pulsed photoactivation. Signal arrival rates to distinct regions equidistant from the photoactivated center show marked disparities (varying by more than a factor of 4 within a single cell). We compare these measurements to simulations of diffusing particles on the visualized ER network structure, and show that the simulated rates of arrival to distinct regions show strong agreement with experimental data. These results demonstrate the importance of network structure in guiding the observed heterogeneity in protein spread.



By modifying and extending a model for `minimal networks' driven by membrane tension and new tubule growth~\cite{lin2014structure}, we assess the effect of ER network rearrangements on protein spread. The substantial separation of timescales between network dynamics and protein diffusivity leads to only a marginal predicted effect of tubule rearrangement on the motion of proteins within the ER. 

Additionally, we simulate pairs of reactive particles diffusing through the ER and demonstrate that the structural heterogeneity of the network gives rise to effective hot spots where encounters are more likely to occur. Intriguingly, visualization and analysis of ERES positions across the peripheral ER indicates that these structures are more likely to be found in diffusively accessible hot-spot regions.



We note that the ER models in this study are intentionally highly simplified, reducing the complex membrane-enclosed geometry of the ER to a network of effectively one-dimensional tubules. These simplifications make it possible to focus on the role of cellular-scale network connectivity and rearrangements in particle transport. Notably, the simple structural model is sufficient to reproduce the observed heterogeneous protein arrival rates to different network regions in photoactivation experiments.
More detailed structural models could include variability in tubule diameter~\cite{zucker2022mechanism,wang2022endoplasmic} as well as scattered peripheral sheets~\cite{shibata2010mechanisms}, which may themselves be perforated with holes~\cite{schroeder2019dynamic} or composed of dense tubular matrices~\cite{nixon2016increased}. 
Exploring the effect of these structures on particle transport forms a potentially interesting avenue for future work.


The network dynamics model employed here aims to isolate the key important features governing ER rearrangements -- namely, the formation of new tubules and the tension-driven movement of junctions~\cite{westrate2015form,upadhyaya2004tension,lin2014structure}. Although network dynamics are shown to have little impact on protein diffusion, they are expected to play a greater role in the motion of larger and slower-moving ER-associated bodies such as the ERES~\cite{stadler2018diffusion}. Furthermore, it is possible that directed flows of luminal and / or membrane contents may be associated with the growth and shrinking of ER tubules, as implied by recent evidence that new tubule growth is followed by a delayed widening and infilling with Climp63 spacer proteins~\cite{wang2022endoplasmic}. Although the spatial extent and magnitude of such flows is not currently established, they could more extensively contribute to modulating intra-ER protein motion.



The transport, quality control and export of proteins in the ER are essential biological processes in the early secretory pathway. These processes require a variety of encounters between newly manufactured proteins, chaperones, and regulatory factors. The structural heterogeneity of the ER network implies that certain regions may allow for more efficient encounters between binding partners. Notably, however, the effect of morphology becomes important only in the regime of diffusion-limited kinetics when the particles are sparsely scattered over the network~\cite{viana2020mitochondrial}. The sequestration of some quality control machinery to specific regions of the ER~\cite{sontag2014sorting} implies that long-range diffusive search by proteins through the network may be an important factor in the kinetics of such pathways.

In addition to protein transport, the results described here apply to any diffusive particles contained in the membrane or lumen of the peripheral ER network. This includes ions such as calcium, as well as the buffer proteins which bind to them. In particular, we would expect the demonstrated structural heterogeneity of the ER to lead to more rapid calcium release in better-connected regions of the network. Given that calcium homeostasis and signaling is one of the key functional roles of the ER, heterogeneous transport could thus provide an important link between physical structure and biological function. Furthermore, it would be interesting to explore whether contacts between the peripheral ER and other cellular structures, such as mitochondria, tend to preferentially occur at highly connected regions, which may facilitate the delivery of lipids or ions across these contacts~\cite{toulmay2011lipid,reane2020er,katona2022capture}.


Through the use of experiments paired with quantitative image analysis and computational modeling, our results demonstrate how morphology guides particle transport and reactions in the ER, with broad implications for diffusive transport in any intracellular network structure.

\section*{Author Contributions}
ZCS, LMW, and EFK conceived and designed the research and wrote the manuscript. ZCS and EFK developed and implemented the model for both particle diffusion and network dynamics. ZCS analyzed imaging and simulation data. KK, MV, LC, and LMW generated experimental data and performed imaging studies. 

\section*{Acknowledgments}
We thank Dr. Edward Avezov and Dr. Eric Arnoys for helpful discussions and the Van Andel Institute Optical Imaging Core (RRID:SCR\_021968), especially Corinne Esquibel for their assistance with the Zeiss LSM 880. Funding was provided by the National Science Foundation (Grant ID \#2034482 to EFK and \#2034486 to LMW), as well as a Cottrell Scholar Award from the Research Corporation for Science Advancement to EFK.

\bibliography{ERheterogeneity}


\end{document}